# A Statistical Approach Towards Robust Progress Estimation

Arnd Christian König
Microsoft Research
One Microsoft Way
Redmond, WA 98052, USA
chrisko@microsoft.com

Bolin Ding
University of Illinois at
Urbana-Champaign
201 N. Goodwin Avenue
Urbana, IL 61801, USA
bding3@uiuc.edu

Surajit Chaudhuri, Vivek
Narasayya
Microsoft Research
One Microsoft Way
Redmond, WA 98052, USA
{surajitc,
viveknar}@microsoft.com

## ABSTRACT

The need for accurate SQL progress estimation in the context of decision support administration has led to a number of techniques proposed for this task. Unfortunately, no single one of these progress estimators behaves robustly across the variety of SQL queries encountered in practice, meaning that each technique performs poorly for a significant fraction of queries. This paper proposes a novel *estimator selection* framework that uses a statistical model to characterize the sets of conditions under which certain estimators outperform others, leading to a significant increase in estimation robustness. The generality of this framework also enables us to add a number of novel "special purpose" estimators which increase accuracy further. Most importantly, the resulting model generalizes well to queries very different from the ones used to train it. We validate our findings using a large number of industrial real-life and benchmark workloads.

## 1. INTRODUCTION

Accurate estimation of the progress of database queries can be crucial to a number of applications such as administration of long-running decision support queries. As a consequence, the problem of estimating the progress of SQL queries has received significant attention in recent years [6, 13, 14, 5, 12, 16, 15, 17]. The key requirement for all of these techniques (aside from small overhead and memory footprint) is their *robustness*, meaning that the estimators need to be accurate across a wide range of queries, parameters and data distributions.

Unfortunately, as was shown in [5], the problem of accurate progress estimation for arbitrary SQL queries is hard in terms of worst-case guarantees: none of the proposed techniques can *guarantee* any but trivial bounds on the accuracy of the estimation (unless some common SQL operators are not allowed). While the work of [5] is theoretical and mainly interested in the worst case, the property that no single proposed estimator is robust in general holds in practice as well.

We find that each of the main estimators proposed in the literature performs poorly relative to the alternative estimators for some (types of) queries. To illustrate this, we compared the estimation errors for 3 major estimators proposed in the literature (DNE [6], the estimator of Luo et al (LUO) [13] and the TGN estimator based on the *Total GetNext* model [6] tracking the GetNext calls at each node in a query plan) over a number of real-life and benchmark workloads (described in detail in Section 6). We use the average *absolute* difference between the estimated progress and true progress as the estimator error for each query and the compare the ratio of this error to the minimum error among all three estimators. The results are shown in Figure 1, where the Y-axis shows the ratio and the X-axis iterates over all queries, ordered by ascending ratio for each estimator – note that the Y-axis is in log-scale. As we can see, each estimator is (close to) optimal for a subset of the queries, but also degrades severely (in comparison to the other two), with an error-ratio of 5x or more for a significant fraction of the workload. No single existing estimator performs sufficiently well across the spectrum of queries and data distributions to rely on it exclusively.

However, the relative errors in Figure 1 also suggest that by judiciously selecting the best among the three estimators, we can reduce the progress estimation error. Hence, in absence of a single estimator that is always accurate, an approach that chooses among them could go a long way towards making progress estimation robust.

Unfortunately, there appears to be no straightforward way to precisely state simple conditions under which one estimator outperforms another. While we know that e.g., the TGN estimator is more sensitive to cardinality estimation errors than DNE, but more robust with regards to variance in the number of GetNext calls issued in response to input tuples, neither of these effects be reliably quantified before a query starts execution. Moreover, a large numbers of other factors such as tuple spills due to memory contention, certain optimizations in the processing of nested iterations (see Section 5.1), etc., all impact which progress estimator performs best for a given query.

### 1.1 Our Approach

Given this complexity, the task of manually formulating a decision function for selecting among progress estimators appears daunting; in fact, there are some hardness results on choosing among estimators (see [5], Section 6.3.) that again indicate that this problem is hard in a worst-case sense. However, we are encouraged by the fact that it is relatively easy to (a) obtain large numbers of examples of the performance of various progress estimators for different queries and data distributions and (b) identify several factors which are weakly predictive of the best estimator to use for a given query. Therefore, leveraging statistical machine learning, which successfully has been applied to many similar scenarios, where the and training data (such as changes in data size, skew, physical design and selectivity) on overall progress estimation quality. We will also show that the techniques have very low overhead, as the computation of all proposed features is very simple; the model used for estimator selection can be (re-)trained very quickly even for very large numbers of example queries, allowing for the fast incorporation of additional training observations if needed.

## 2. RELATED WORK

The problem of progress estimation for database queries was first proposed and studied in [6] and [13], with further refinements proposed in [5, 14]. We will describe these techniques in detail in Section 3.4 and use them as candidate progress estimators which our technique then selects from. [12] extends the earlier work to multi-query progress estimation, and [17] studies progress estimation in MapReduce clusters, both of which are scenarios that are beyond the scope of this paper, but are important extensions for future work.

The work of [16, 15] introduces techniques that improve the optimizer estimates of join and group sizes (which are then used in progress estimation) by taking random samples of the underlying data distributions online for operator-pipelines that involve a 'preprocessing phase' (such as the build phase during a hash join). This can provide more accurate estimates with probabilistic accuracy guarantees; however, the processing of the samples can induce significant increases in run-time (e.g., see Table III in [15] where the authors report run-time increases of 5% for larger scale factors). Given the desire for low overhead of progress estimation in large-scale systems, this overhead is the main reason why we do not include this technique in our experimental comparison in Section 6. For operators without a natural preprocessing stage (such as nested loop joins) the authors of [15] propose using a random sample of the outer relation which is stored as a "prefix" of the underlying table (i.e., it is read first and joined with the inner relation to arrive at a join cardinality estimate). Here, it is not clear how to maintain these prefixes under insertions/deletions to the base tables or for intermediate query results.

Statistical (or machine-learning based) models have been proposed for the task of modeling SQL query a number of times before (e.g., [11, 8]); both of these techniques do not address progress estimation itself, but the related problem of estimating the duration of query execution. Both of these approaches model the target workload directly (as opposed to the two-step approach of this paper) and thus have a very strong dependence of the observations used to train the model being very similar to the one the model is deployed on. For example, when the training and test queries used different schemata and databases, the approach of [11] often resulted in predicted run-times that were up to multiple orders of magnitude longer than the actual time the query ran (see Experiment 4, Figure 15 in [11]).

The implicit assumptions in [8] are even more stringent: each query (template) seen in the "test" data is seen during the training phase as well (often, multiple times) and each query is associated with a single plan (the latter restriction can be lifted, however, by treating different plans as different queries). Under these assumptions, generalizing to "ad-hoc" queries is not an option.

## 3. BACKGROUND

In this section we first introduce some required notation, describe the notions of query *pipelines* and the techniques used to refine optimizer estimates in the context of progress estimation and give an overview of existing progress estimators.

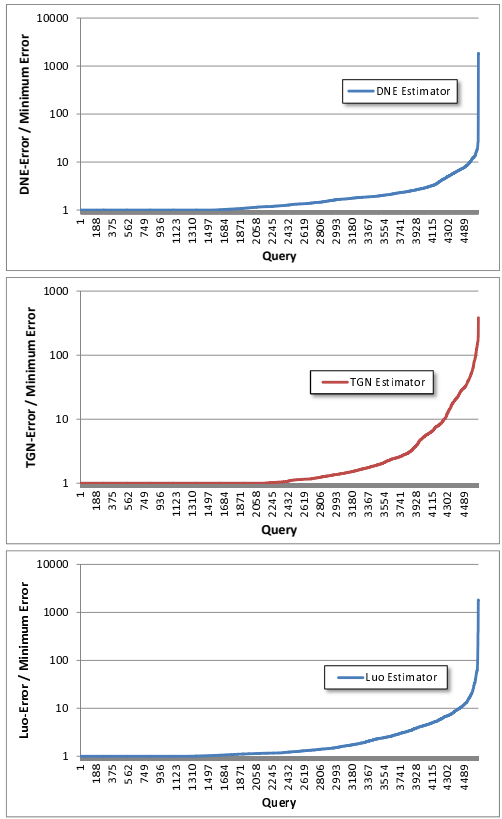

**Figure 1: Each of the 3 progress estimators performs poorly (relative to the best among the others) for a large subset of queries.**

complexity and number of input "features" made manual reasoning infeasible (such as text categorization [18] or web-scale information retrieval [4]), but abundance of training data allowed for automated inference of decision functions, appears promising. In this paper, we will attempt to define such a statistical model for the task of selecting among a set of progress estimators the one with the lowest error, and identify features that are predictive of the correct choice and that can be obtained within a DBMS at very low overhead.

While statistical machine learning models are known to be effective when there is a strong similarity between the data the model is trained on and the data it is used for, we cannot assume this to be the case when deploying the learned model in production. Hence, a key challenge to this approach comes from the fact that we want to do well for any "ad-hoc" queries, i.e., we must ensure that when we are faced with different queries (or different data sets) than the query examples used to train the model, we continue to perform well in the vast majority of instances. This means that we need to identify features to use in the model that generalize well across very different queries, and evaluate our technique's ability to remain accurate when faced with completely new workloads.

As we will show, our approach does indeed generalize well across workloads, even on different datasets, and leads to substantial reduction in the progress estimation error when compared to any of the previously proposed progress estimators. In order to demonstrate this, this paper will focus on a very detailed empirical evaluation, for which we will use a large set of different queries from both synthetic and real-life query workloads and data distributions and study the effects of various types of differences between test

## 3.1 Notation

The earlier progress estimation techniques of [6, 13, 14, 5, 16, 15] all model a running query $\mathcal{Q}$ as a tree of physical operators; we use $Nodes(\mathcal{Q}) \subset \mathbb{N}$ to enumerate the nodes in the plan and $Op(i), i \in Nodes(\mathcal{Q})$ to denote the physical operator at a node $i$. We denote the set of all nodes below $i$ in the execution plan as $Decendants(i)$.

All known progress estimation techniques base their estimates on a set of counters collected at various nodes in the physical plan of a query, which are observed at various points during the query's execution. We denote the set of all such observation of a query $\mathcal{Q}$ as $Observations(\mathcal{Q})$ and – when it is not clear from context – we use a variable $t \in Observations(\mathcal{Q})$ to specify at which point in time a specific counter-value was observed. Each observation is associated with the time $Time(t)$ at which it occurred; we also use $t_{start}$ and $t_{end}$ to denote the first/last observation made, corresponding to the start/termination of a query.

The following counters are used by the different techniques to estimate the progress of $\mathcal{Q}$:
(1) The total number of *GetNext* calls at node $i$ issued at point $t$ in the query's execution; we denote this value by $\mathcal{Q}.K_i^t$; when $\mathcal{Q}$ or $t$ are clear from context, we drop them from the notation. Note that disk spills due to lack of memory (e.g., in hybrid hash joins) are modeled as additional *GetNext* calls as well.
(2) The total number of *GetNext* calls at node $i$ issued over the *entire* duration of $\mathcal{Q}$ (denoted by $\mathcal{Q}.N_i$). Note that the value of this counter is generally not known for all nodes in a plan before the query terminates.
(3) The current estimate of the total number of *GetNext* calls at node $i$ issued over the duration of $\mathcal{Q}$ (denoted by $\mathcal{Q}.E_i^t$). Since the value of $N_i$ is typically not known during the execution of $\mathcal{Q}$, most progress estimators use this estimate instead. The value of $E_i$ is generally derived from the optimizer estimate at the beginning and then progressively refined as the query executes (which we will describe in detail in Section 3.3).
(4) Absolute bounds on the value of $N_i$ which are based on the cardinality of the inputs to a node and refined as the query executes; here, we track both lower ($\mathcal{Q}.LB_i^t$) and upper ($\mathcal{Q}.UB_i^t$) bounds.
(5) The number of bytes (logically) read ($\mathcal{Q}.R_i^t$) and written ($\mathcal{Q}.W_i^t$) at a node.

## 3.2 Pipelines/Segments

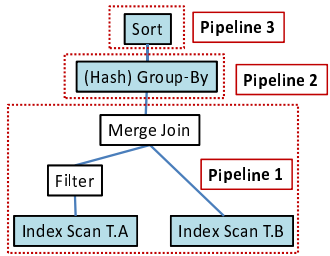

**Figure 2: Example execution plan with multiple pipelines.**

To capture the notion of nodes in a query plan which execute concurrently, prior work defined the notion of *Pipelines* or *Segments* (defined in [6], and [13], respectively), which correspond to maximal subtrees of consecutively executing nodes in a (physical) operator tree. We denote the set of all pipelines/segments for a query $\mathcal{Q}$ as $Pipelines(\mathcal{Q}) = \{P_1, \ldots, P_l\}$; each pipeline $P_j$ is defined via a subset of nodes $P_j \subseteq Nodes(\mathcal{Q})$.

For each pipeline (or segment), the nodes that are the sources of tuples operated upon by the remaining nodes (i.e., typically all leaf nodes of the pipeline, excluding the inner subtree of nested loop operators) are referred to as the *driver nodes* (or *dominant input*) of the pipeline. We use the set $DNodes(P_j) \subseteq P_j$ to denote the driver nodes in pipeline $P_j$. An example operator tree with 3 pipelines is shown in Figure 2; the shaded nodes correspond to driver nodes. The notion of segments was further refined in [14].

## 3.3 Refining Cardinalities

Accurate estimates of the number of tuples at different operators is of crucial importance to accurate progress estimation. However, unlike e.g., query optimization, progress estimation can easily leverage improved estimates obtained *after* a query has started to execute (whereas e.g., changing query plans mid-flight is a non-trivial operation). As a consequence, all progress estimation techniques incorporate some form of online refinement of optimizer estimates.

The techniques in [6] maintain *worst-case* bounds on the number of GetNext calls, based on the number of tuples seen so far and the size of the input(s). If the value of any $E_i$ ever falls outside of these bounds, it is set to the nearest boundary value.

[13] uses a more aggressive strategy, updating the $E_i$ counters using interpolation: for any pipeline, this approach first measures the percentage of the dominant input(s) consumed in a pipeline $p_j$ as

$$\alpha_{P_j} = \frac{\sum_{i \in DNodes((P_j))} K_i}{\sum_{i \in DNodes((P_j))} E_i}. \quad (1)$$

Now, given that a node has output $K_l$ tuples so far, we can extrapolate the number of tuple output by this node as $\hat{E}_l = K_l / \alpha_{P_j}$; the refinement technique in [13] then interpolates between this value and the initial estimate $E_l$ as follows, reflecting the growing confidence in $\hat{E}_l$ as the query progresses:

$$E_l^{new} = \alpha_{P_j} \times \hat{E}_l + (1 - \alpha_{P_j}) \times E_l. \quad (2)$$

Finally, the techniques of [15] leverage random samples of (some of) the input relations to give more accurate estimates during query execution. The use of random samples enables them to provide probabilistic guarantees of the accuracy of these estimates, but it requires to either randomly sample from the tuple stream during query execution (which is potentially expensive) or to pre-compute these (which is also costly for update-heavy data and not always be possible in practice).

## 3.4 Previous Estimators

In the following we briefly review the progress estimators proposed previously. [6] introduced the concept of progress estimation based on the so-called *GetNext Model* of progress, which assumes that the total work (i.e., CPU overhead, I/O, etc.) is amortized across the *GetNext* calls issued across *all* nodes in the execution plan. Hence, the fraction of the total *GetNext* calls executed at any point in a query can be used as an estimate of its progress.

Unfortunately, the total number of such calls ($\sum N_i$) issued over the course of a query's execution is generally not known before the query is completed, meaning that the value of this model is primarily theoretical. Instead, the corresponding optimizer estimates $E_i$ have to be used in place of $N_i$, resulting in the following estimator:

$$\text{TGN}_\mathcal{Q} = \frac{\sum_{i \in Nodes(\mathcal{Q})} K_i}{\sum_{i \in Nodes(\mathcal{Q})} E_i}. \quad (3)$$

Note that the accuracy of this estimator critically depends on the accuracy with which the $E_i$ values are estimated. However, note that in many cases the exact sizes of the inputs to the *driver nodes* of a pipeline are known at the beginning of the pipeline's execution. Now, under the assumption that the number of *GetNext* calls occurring in the pipeline overall is proportional to the number of input tuples read from the driver nodes, it is possible to (somewhat) sidestep the issue of inaccurate estimation and propose the following *DriverNode* estimator for the progress of a pipeline:

$$\text{DNE}_{P_j} = \frac{\sum_{i \in DNodes((P_j))} K_i}{\sum_{i \in DNodes((P_j))} E_i}. \quad (4)$$

This estimator performs well in scenarios where the amount of GetNext calls triggered by (equal-sized groups of) input tuples is roughly equal, but performs less well when there is significant variance in this "per-tuple work".

The progress of the entire query is now the weighted sum of the pipelines' estimated progress:

$$\text{DNE}_{\mathcal{Q}} = \sum_{P_j \in Pipelines(\mathcal{Q})} \frac{\text{DNE}_{P_j} \times \sum_{i \in DNodes((P_j))} E_i}{\sum_{i \in Nodes(\mathcal{Q})} E_i}. \quad (5)$$

The estimators proposed in [13] also leverage the notion of pipelines (called *segments*) and driver nodes (*dominant input(s)*), but only track the bytes read at the pipeline inputs and written at the pipeline output (in addition to reads/writes of spills). We shall refer to this as the *Bytes Processed Model* of progress. Moreover, the approach converts this measure into an absolute run-time (as opposed to a percentage) by measuring the speed at which these bytes are read/written during the last $T$ (typically set to 10) seconds. We refer to the corresponding progress estimator as $\text{LUO}_{\mathcal{Q}}$.

The paper [5] introduced two additional estimators, PMAX and SAFE. It can be shown the *ratio error* for the PMAX estimator (when compared to the GetNext model using completely accurate cardinality estimates) is within a factor of $\mu$ of the correct progress (where $\mu$ is average number of *GetNext* calls performed during the entire query per input tuple), whereas the SAFE estimator is *worst-case optimal* (again, with respect to the ratio error). Note that while the ratio error is of significant theoretical interest, it is unlikely the most relevant error metric for practical use, as it overemphasizes errors very early in a query's execution; instead, most (but not all) of our experiments will focus on absolute errors.

## 4. ESTIMATOR SELECTION

In this section we will describe the *Estimator Selection* module that decides which estimator among a given set of candidate estimators to use for progress estimation. The overall architecture is shown in Figure 3. For any incoming query $\mathcal{Q}$ the execution plan is first generated by the optimizer (or matched in the plan cache); from this plan we obtain the estimated number of GetNext calls $E_i$ (and – by multiplying these with the average row width – the estimated number of bytes processed required for the LUO estimator). As input to the estimator selection, we use two types of features: the first set are the so-called *static features* (whose values are determined *before* the query starts execution) which are based on the shape of the execution plan and optimizer estimates. We describe these in detail in Section 4.3. These features are now used by the estimator selection module to select the progress estimator with the smallest estimation error among the techniques described in Section 3.4 (and a number of new progress estimators we will introduce in Section 5). In particular, as progress estimators for different pipelines are independent and can be combined as a weighted sum (see Section 3.4), we can select (potentially different) estimators for each pipeline/segment in the execution plan.

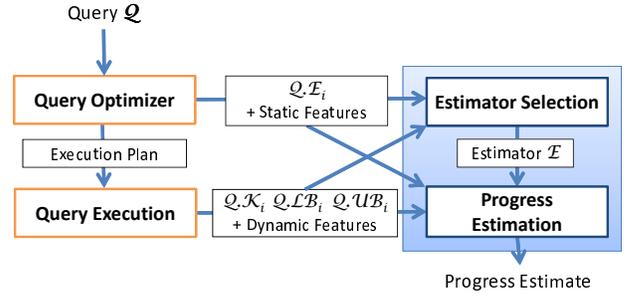

**Figure 3:** Overview of our solution.

Now, as $\mathcal{Q}$ starts execution, we obtain additional counters from the query execution engine which are used by the chosen progress estimator(s) to produces an estimate of the overall query progress. Moreover, observing the query execution allows us to use additional properties of the query (such as the observed variance in the number of GetNext calls triggered by tuples read from the Driver Nodes) which we were not able to assess before. Consequently, we use this data to formulate so-called *dynamic features* which are also input to the estimator selection and allow us to – when necessary – to revise the initial estimator choices online.

### 4.1 The Learning Task

Next, we will describe the way we model estimator selection as a machine learning task. Given that our goal is to, when given an input query, select one among a fixed set of estimators, a natural setup appears to formulate this as a multi-class classification problem. However, the key quantity we seek to optimize is not only the estimator choice, but the size of the overall progress estimation error. In particular – for many combinations of query and estimator – many estimators do produce (almost) identical estimates (and thus errors), meaning that there can be several optimal choices. We need to differentiate such cases from those in which the difference in estimator accuracy is very large; thus, it is crucial to identify and model the cases where a given estimator performs very poorly, thereby allowing the technique to minimize the expected impact of selection errors, something that a simple classification set-up does not allow us to do.

As a consequence, we do not model estimator selection as a classification (or learning-to-rank) task. Instead, we propose a setup where – for each candidate estimator – we train a regression model that models the estimation error when using the estimator in question. The estimator selection module then selects the estimator with the smallest predicted error.

It is important to note that this learning task is significantly different from (and likely simpler than) the setup of [11] where the explicit runtime of $\mathcal{Q}$ is predicted *before* execution. For one, our models can leverage fairly simple structural properties of pipelines (e.g., the presence of nested loop joins, which cause issues for DNE estimators), which by themselves do not give any direct hints regarding absolute runtime. Second, our setup offers various forms of online adaptivity such as the online cardinality refinement described in Section 3.3 and the dynamic features obtained as the query is executing, allowing us to improve initial estimates.

**Combining Estimators:** In addition to this framework which selects a single estimator, we also experimented with models that estimate progress as the weighted combination of estimators. Here, we found that a simple approach that uses training data to compute static weights for each different estimator did not perform well. The different weights varied considerably with the relative frequencies of different types of queries in the training workload (e.g., the relative weight assigned to the DNE estimator would fluctuate with the number of queries with nested loop joins in the training set). As a result, the resulting combined estimator was not robust in the sense that – if the queries it was used for were very different from the training set – the resulting estimation errors were often very high. We did not pursue this type of approach further.

### 4.2 The Learning Model

The learning method we use to compute regression estimates of the individual estimator's errors is based on *Multiple Additive Regression-Trees* (*MART*). MART is based on the *Stochastic Gradient Boosting* paradigm described in [10] which performs gradient descent optimization in the functional space. In our experiments, we used the *(root) mean square error* as the loss function (optimization criterion), used the steepest-decent (gradient descent) as the optimization technique, and used binary decision trees as the fitting function - a "nonparametric" approach that applies numerical optimization in functional space.

In an iterative boosting (or residue-learning) paradigm, at the beginning of every iteration, the estimation errors of the training data are computed using the current model. The error prediction is then compared with the actual error outcome to derive the errors (or residuals) for the current system, which is then used to fit a residue model – a function that approximates the errors – using *MSE* (Mean Square Error) criteria. In MART, we compute the derivatives of the log-loss for each training data point as the residual and use the regression tree as the approximation function-residual model. A regression tree is a binary decision tree, where each internal node splits the features space into two by comparing the value of a chosen feature with a pre-computed threshold; once a terminal node is reached, an optimal regression value is returned for all the data falling into the region. Finally, the residual model is added back to the existing model so that the overall training error is compensated for and reduced for this iteration. The new model – the current plus the residual model – will be used as the current model for the next boosting/training iteration. The final model after $M$ boosting iterations is the sum of all $M$ regression trees built. A a more detailed description of MART can be found in [19].

**MART vs. other statistical models:** In addition to MART, we initially also evaluated our approach using different statistical models (including logistic regression models and Support Vector Machines). Here we found that using MART resulted in significantly better accuracy, which appears to be in part due to two properties of MART: for one, MART does not require transformations to normalize the inputs into zero mean and unit variance which is essential for other algorithms such as logistic regression or neural nets. More importantly, by its internal use of decision trees, which are able to "break" the domain of each feature arbitrarily, MART is able to handle the non-linear dependencies between the feature values and the estimation errors without using explicit *binning* as a preprocessing step. This allows us to use a much wider range of features than linear models would be able to leverage.

### 4.3 Static Features

We will now define the *static features* used by our technique, i.e., the set of features that do not depend on feedback from query execution itself. First, we need to encode the physical execution plan for a pipeline $P_j$ for which we want to select an progress estimator. One encoding that has been proposed in the context of execution time prediction is the one of [11], which uses – for each physical operator type *op* known to the execution engine – the number of instances of *op* in the plan

$$Count_{op} = |\{i \in Nodes(P_j) : Operator(i) = op\}|$$

and the cardinalities of tuples for each such operator type

$$Card_{op} = \sum_{\substack{i \in Nodes(P_j) \\ Operator(i) = op}} E_i \text{ as features.}$$

However, this encoding uses the *absolute* cardinalities, which are more relevant in the context of execution time prediction, but can be misleading in the context of progress estimation: for example, when a nested loop join is expected to output 100K tuples, this might be negligible compared to the outer input being several millions of tuples (thereby making a DNE estimator a likely candidate), but very significant when the outer input is very small. Therefore, in addition to the above features, we also encode the *relative* cardinalities as well. In particular, we encode the cardinalities (relative to the (estimated) total number of tuples) *at* nodes with an operator *op*, as well as the relative cardinalities of their "input subtrees" and of the subtrees they feed into, formally defined as follows:

$$SelAt_{op} = \frac{\sum_{i \in Nodes(P_j) \wedge Operator(i)=op} E_i}{\sum_{i \in Nodes(P_j)} E_i}$$

$$SelAbove_{op} = \frac{\sum_{i \in Nodes(P_j) | \exists j: Operator(j)=op \wedge j \in Decendants(i)} E_i}{\sum_{i \in Nodes(P_j)} E_i}$$

$$SelBelow_{op} = \frac{\sum_{i \in Nodes(P_j) | \exists j: Operator(j)=op \wedge i \in Decendants(j)} E_i}{\sum_{i \in Nodes(P_j)} E_i}$$

To illustrate these definitions, consider the query plan in Figure 2; here, the value of $SelBelow_{Merge\_Join}$ would include the $E_i$ values of the *filter* node and the two *index scans*, whereas $SelBelow_{Filter}$ would include the $E_i$ value at the merge join. Finally, we also encode the cardinalities (relative to the (estimated) total number of tuples) of all driver nodes in a pipeline in the feature $SelAt_{DN}$.

### 4.4 Dynamic Features

In the following we will describe the *dynamic* features (i.e., features that are obtained *after* a query has started execution) we use for estimator selection. Using these we may revise our initial choice of progress estimators as a query executes, ideally arriving at a more accurate estimator as the query goes on. Given that for applications of progress estimation such as deciding whether to cancel a long-running query the importance of accurate estimates becomes larger as the query executes, being able to leverage this execution feedback for estimator selection can be quite valuable. In the following, we will first illustrate factors in estimator selection that can not be captured well using static features, but can be assessed more accurately after a query has started executing. We then describe the dynamic features we use in detail.

#### 4.4.1 Limitations of Static Features

Given that the set of static features defined above already exceeds the ones proposed in [11], one might question why additional features are necessary. To illustrate this, we will first demonstrate that the estimation errors for two simple progress estimators (DNE and TGN) are a function of (a) the "variance" in the amount of work done for the tuples from the driver node input (over the different observations) and (b) the cardinality estimation error(s) made by the optimizer. Both of these cannot be reliably assessed before query execution. However, for many pipelines, we can estimate both of these factors with some degree of accuracy by monitoring the flow of tuples during the pipeline's execution. In fact, under the assumption of (nearly) random tuple orderings (and limited skew), the initial estimates of both of these factors become rather accurate after only a small fraction of the query has executed, which we can then use to revise the initial choice of progress estimators. Note that while the worst-case results of [5] with regards to determining the most accurate estimator online continue to hold, these rely on specific data distributions with adversarially chosen data ordering, which are highly unlikely in practice.

First, consider the estimation error for the DNE estimator at an observation $t$. For the purposes of this example, we will assume that the *GetNext Model* (which corresponds to the TGN estimator using the correct $N_i$ in place of their estimates $E_i$) is the correct "gold standard" of progress (how well this assumption holds in practice is something we will experimentally verify in Section 6.7). We will consider a trivial execution plan, containing two nodes, a *table scan* (node 0) followed by a *filter* (node 1), with node 0 being the driver node.

Given this setup, the average number of GetNext calls triggered by each tuple read at the driver node is $N_0/N_1$. Now, the absolute progress estimation error for an observation $t$ can be written as:

$$\left\| \overbrace{\frac{K_0^t}{E_0^t}}^{\text{DNE Estimator}} - \overbrace{\frac{K_0^t + K_1^t}{N_0 + N_1}}^{\text{"True" Progress}} \right\| = \quad [\text{ as driver-node input size is known }]$$

$$\left\| \frac{K_0^t}{N_0} - \frac{K_0^t + K_1^t}{N_0 + N_1} \right\| = \left\| \frac{\overbrace{\left((N_1/N_0) \cdot K_0^t - K_1^t\right)}^{\text{"Variance" in GetNext calls}}}{N_1 + N_0} \right\|$$

Thus, the progress estimation error corresponds to the (weighted) difference between the expected work at node 1 ($= (N_1/N_0)K_0^t$) – and the actually observed one ($= K_1$). Similar constructions also hold for more complex plans. Using a similar model construction for the TGN estimator, we obtain:

$$\left\| \overbrace{\frac{K_0^t + K_1^t}{E_0^t + E_1^t}}^{\text{TGN Estimator}} - \overbrace{\frac{K_0^t + K_1^t}{N_0 + N_1}}^{\text{"True" Progress}} \right\| = \quad [\text{ as driver-node input size is known }]$$

$$\left\| \frac{K_0^t + K_1^t}{N_0^t + E_1^t} - \frac{K_0^t + K_1^t}{N_0 + N_1} \right\| = \left\| \frac{K_0^t + K_1^t}{(N_0+N_1)(N_0+E_1^t)} \overbrace{(N_1 - E_1^t)}^{\text{Estimation Error}} \right\|$$

Thus, here the progress estimation error is a weighted function of the error in the query optimizer's cardinality estimation for node 1. Again, similar constructions also hold for more complex plans.

Now, since both the variance in per-tuple work and the cardinality estimation error can (for many combinations of queries and data orderings) be estimated with some degree of accuracy online when observing the execution of a query (using the same counters that progress estimators are based on), this motivates the use of "dynamic" features which are computed during (the early stages of) query execution and which indirectly leverage information about both factors: while the dynamic features we propose do not quantify the variance in "per-tuple work" or cardinality estimation errors directly, they seek to characterize their effects on the various progress estimators with regard to the *Total GetNext Model* as well as the actual time passed between observations.

### 4.4.2 Defining Dynamic Features

The first set of dynamic features we will describe quantify the relative differences in estimation provided by different progress estimators at different points in a query's execution. To illustrate how such features could potentially be useful, consider the case where we measure the difference between the progress estimate provided using the DNE and TGN estimators for a nested-loop join pipeline. Now, if both progress estimators make similar progress during the initial part of the query, but the TGN estimator subsequently makes much more rapid progress, this can be an indicator of some tuples from the driver node joining with a large number of tuples on the inner side of the join, which in turn indicating large variance in "per input-tuple work".

Ideally, we would want to compute these differences between estimators at consistent points in query execution (so that we have similar points of reference when training the estimator selection model), e.g., at 2%, 5% and 10% of a query's execution. Obviously, this is not feasible – if we knew during a query's execution which fraction of it was done, progress estimation would be trivial. Instead, we use the fraction of the driver node input (the size of which is know up-front for many queries) that has been consumed to create these "markers" when generating a feature. For this purpose, we first define $t\{x\} \in Observations(\mathcal{Q})$ as the first observation for which $x\%$ of the driver node input was "consumed", i.e., $\sum_{i \in DNodes} K_i^t / \sum_{i \in DNodes} E_i^t \geq x/100$. We also use the notation $\text{DNE}^{t\{x\}}$ to denote the DNE estimate of progress at observation $t\{x\}$ (and similarly for other progress estimators). Now, we define the difference between DNE and TGN "at $x\%$" as

$$\text{DNE}vs\text{TGN}_x = \left\| \text{DNE}^{t\{x\}} - \text{TGN}^{t\{x\}} \right\|.$$

We then use these differences – for $x \in \{1, 2, 5, 10, 20\}$ – as features in estimator selection. Similarly, we define the differences between all other estimators, including the novel estimators which we will introduce in Section 5.

The second set of dynamic features quantifies how well the estimators correlate with time itself over the fraction of the query we have observed so far; for this purpose, we take a sequence of $k$ observations $t\{x/k\}, t\{(2x)/k\}, \ldots, t\{(kx)/k\}$ and compute – for each observation – the fraction of time since the start of the query and the fraction of progress that – according to different estimators – these points correspond to.

For example, for the DNE estimator, we compute the following features for $i = 1, \ldots, k$:

$$Cor_{\text{DNE}, i, x} = \frac{Time(t\{ix/k\}) - Time(t_{start})}{Time(t\{x/k\}) - Time(t_{start})} \cdot \frac{1}{\text{DNE}^{t\{x\}}}$$

## 5. NOVEL ESTIMATORS

Given the framework for estimator selection introduced in the previous section, we can now also consider extending it with additional progress estimators. As long as we can rely on the framework to pick a suitable estimator (with high likelihood), these progress estimators do not necessarily need to be well-suited for arbitrary queries/pipelines, but only have to improve progress estimates for specific scenarios. Here, we propose the different estimators to address the effects of different issues such as correction of selectivity errors, dealing with "partial" batches, etc. on progress estimation in practice.

### 5.1 Estimators for Batch Operations

One assumption implicit in most estimators discussed so far is that – within a pipeline or segment – tuples flow from the input driver nodes to the top node of the pipeline (or are filtered) without any blocking. However, there exist some query processing optimizations that do result in blocking effects. For example, one common optimization used for nested iterations is the introduction of "partial" batch sorts. Here, a portion of the outer input is sorted in order to subsequently localize references in the inner subtree (for details see e.g., [9]), which can yield significant improvements in nested loop join processing [7].

This optimization may lead to significant inaccuracy for estimators that base the overall progress significantly on the progress seen at the diver nodes (dominant inputs), especially for larger batch sizes, which e.g., may occur for longer-running queries for which some query execution engines increase batch sizes dynamically (e.g., see [9], Section 8.3.). In order to deal with these types of queries, we propose a variant of the DNE estimator which includes these types of batch sort operators among the driver nodes:

$$\text{BATCHDNE}_{P_j} = \frac{\sum_{i \in DNodes(P_j) \vee Op(i)=BatchSort} K_i}{\sum_{i \in DNodes(P_j) \vee Op(i)=BatchSort} E_i} \quad (6)$$

### 5.1.1 DNE with IndexSeeks

As discussed in [5] in detail, a significant challenge to progress estimation and in particular estimators based on the progress seen at driver nodes only are query plans containing nested iterators: when the data distributions on the inner side of the iterator are skewed, the amount of processing and I/O required for tuples from the outer side may vary significantly, which may not be reflected in the estimator. As a consequence, we propose an estimator that behaves like DNE, but also includes all *Index Seek* operations among the driver nodes:

$$\text{DNESEEK}_{P_j} = \frac{\sum_{i \in DNodes(P_j) \vee Op(i)=IndexSeek} K_i}{\sum_{i \in DNodes(P_j) \vee Op(i)=IndexSeek} E_i} \quad (7)$$

## 5.2 Cardinality Interpolation

For this estimator, we adopt the strategy of [13] (detailed in Section 3.3) to refine the estimated number of GetNext calls by interpolating between the initial estimate and the (scaled-up) number of GetNext calls seen so far, and using this new estimate in the TGN estimator. Using $\alpha_{P_j}$ as defined in equation (1) as an estimate of the fraction of the pipeline done so far gives us the following progress estimator:

$$\text{TGNINT}_{P_j} = \frac{\sum_{i \in Nodes(P_j)} K_i}{\sum_{i \in Nodes(P_j)} K_i + (1 - \text{DNE}_{P_j}) \sum_{i \in Nodes(P_j)} E_i} \quad (8)$$

## 6. EXPERIMENTAL EVALUATION

In this section we evaluate the effectiveness of the ideas proposed in this paper and test whether the *estimator selection* technique described in Section 4 significantly improves accuracy of progress estimation. Our evaluation methodology aims to uncover whether the estimator selection method is able to generalize gracefully even as the similarity between the queries used to train the machine learning model and queries used to test the model decreases. We therefore approach this empirical evaluation from three angles. First, we systematically vary key parameters such as data size, physical database design and selectivity that can affect accuracy, and measure how well our estimator selection module is able to select among three existing estimators: DNE, LUO, and TGN. For this controlled experiment, we use the well-known TPC-H decision support benchmark [2] workload over databases with various data distributions. Second, we study the effectiveness of estimator selection for the case of "ad-hoc" queries, i.e. queries that have not executed previously on the system. We study this by training estimator selection on a set of workloads and testing it on an entirely different workload on a different database altogether. We use the TPC-H and TPC-DS decision support benchmarks and two real world decision support workloads (described below).

We complement the above set of experiments on estimator selection robustness by reporting which features of the models were most important (Section 6.5), and measuring the scalability of model training as the size of the training workload increases (Section 6.4). After this, we analyze how many of the proposed progress estimators are really necessary in the context of estimator selection. We conclude this section with a brief discussion of the empirical effectiveness of the *GetNext* and *Bytes Processed* models of progress estimation, which form the theoretical basis of the previous state-of-the-art progress estimators (Section 6.7).

**Experimental Setup:** All experiments were conducted on a 2.66 GHz Intel Xeon PC with 8GB of main memory using a SQL Server 2008 database engine.

**Databases and Workloads:** Since the difficulty of progress estimation for a given query depends on the specific operators present in the query's execution plan as well as parameters such as data size, skew and physical design, we used the following 6 databases and workloads to ensure that we cover a significant range over these parameters:

(1) Over 200 randomly chosen queries from the TPC-DS benchmark [2]. The database size is approx. 10GB. Workloads (2)-(4) each consist of 1000 queries from the TPC-H benchmark [2] – the database is generated using a Zipfian skew-factor Z=1 [1], to induce variance in the "per-tuple work". Because the accuracy of progress estimation can be heavily influenced by the types of execution plans seen (e.g., nested loop joins are known to be more difficult than "*scan based*" queries [5] and typically depend on the existence of indexes on the inner relation), we varied the underlying physical design of the database for the three instances of TPC-H. We considered the following configurations: (a) "*untuned*" which contains only the minimal set of indexes required by integrity constraints, (b) "*fully tuned*" which contains all indexes recommended for this workload by the SQL Server Database Tuning Advisor [3] (DTA) and (c) "*partially tuned*" which contains all indexes recommended by DTA when the space for indexes is restricted to half the space used by the *fully tuned* configuration. To illustrate that this tuning indeed has significant effect on the operators seen during progress estimation, we report the fraction of pipelines containing different operators for TPC-H under the three physical designs in Figure 1. (5) "*Real-1*" is a real-world decision-support and report-ing workload over a 9GB Sales database. Most of the queries in this workload involve joins of 5-8 tables as well as nested sub-queries. The workload contains 477 distinct queries. (6) "*Real-2*" is a different real-life decision-support workload on a larger data set (12GB) with even more complex queries (with a typical query involving 12 joins). This workload contains a total of 632 queries.

| Operator | not tuned | "partially" tuned | fully tuned |
|---|---|---|---|
| NEST. LOOP JOIN | 32.6% | 26.6% | 42.1% |
| MERGE JOIN | 22.7% | 12.8% | 12.9% |
| HASH JOIN/AGG. | 78.8% | 82.9% | 72.9% |
| INDEX SEEK | 47.4% | 65.3% | 96.2% |
| BATCHSORT | 11.7% | 8.3% | 33.9% |
| STREAMAGG. | 18.2% | 9.7% | 21.4% |

**Table 1: Fraction of pipelines with different operators for TPC-H under different physical designs.**

Note that all queries in the different workloads were executed in isolation; extending progress estimation to account for interactions between concurrently executing queries is an interesting challenge, but beyond the scope of this paper.

| Test Set: | "small" queries | "medium" queries | "large" queries |
|---|---|---|---|
| **Estimator** | % optimal | % optimal | % optimal |
| DNE | 18.8% | 29.7% | 44.4% |
| TGN | 71.3% | 51.1% | 40.0% |
| LUO | 9.8% | 19.0% | 15.5% |
| EST. SEL. | 63.9% | 52.0% | 69.0% |

**Table 2: Sensitivity Analysis: varying the total number of GetNext calls between test/training sets. The accuracy of estimator selection does not appear to be affected.**

| Test Set: | "fully" tuned | "partially" tuned | "untuned" |
|---|---|---|---|
| **Estimator** | % optimal | % optimal | % optimal |
| DNE | 30.5% | 40.2% | 44.0% |
| TGN | 41.9% | 26.8% | 38.0% |
| LUO | 27.5% | 33.0% | 18.9% |
| EST. SEL. | 63.9% | 74.7% | 67.6% |

**Table 3: Sensitivity Analysis: varying the physical design between test/training sets. The different physical designs in turn result in different query plans, affecting the choice of optimal estimator.**

| Test Set: | Skew $Z=0$ | Skew $Z=1$ | Skew $Z=2$ |
|---|---|---|---|
| **Estimator** | % optimal | % optimal | % optimal |
| DNE | 35.9% | 54.3% | 49.8% |
| TGN | 52.7% | 28.7% | 32.4% |
| LUO | 11.3% | 17.0% | 17.8% |
| EST. SEL. | 52.7% | 60.0% | 59.5% |

**Table 4: Sensitivity Analysis: varying the data skew between test/training sets.**

**Training Parameters:** For all experiments, we trained the underlying MART models (see Section 4.2) using the $M = 200$ boosting iterations; each decision tree has 30 leaf nodes.

**Dynamic Features:** Among the dynamic features we have defined in Section 4.4, we use the following "pairwise difference" features – DNE$vs$TGN$_x$, DNE$vs$TGNINT$_x$ and TGN$vs$TGNINT$_x$ – as well as these time-correlation features: $Cor_{\text{DNE},i,x}$, $Cor_{\text{TGN},i,x}$, $Cor_{\text{LUO},i,x}$, $Cor_{\text{BATCHDNE},i,x}$, $Cor_{\text{DNESEEK},i,x}$ and $Cor_{\text{TGNINT},i,x}$ (for $i = 1, \ldots, 4$). We use $x \in \{1, 2, 5, 10, 20\}$. This means that in these experiments we stop refining the estimator after 20% of the driver node input has been read. The numbers we report for dynamic features are based on these features only; obviously, further improvements are possible when dynamic features computed at later stages of execution are used – but the resulting improvements are of less consequence as more of the query has executed by the time the improves estimates are seen.

**Error Metric:** Earlier papers on progress estimation have concentrated on both the ratio error as well as the (average) absolute ($L_1$) error between the estimated and true progress. While the ratio error is of significant theoretical interest, it overemphasizes errors that occur very early in the query execution (as either the true progress or the estimated progress will be very small). Instead, our experiments will mainly focus on the absolute difference between the estimated and true progress, using both the $L_1$ and – to further penalize points of large deviation – the $L_2$ norms of these differences. Here, we define the absolute error as the difference between the actual progress of the query (measured based on its overall execution time) and the estimated progress over all observations; for example, the $L_p$ error for the DNE estimator is defined as:

$$\left( \sum_{t \in Observations(Q)} \| (\text{DNE}_Q^t - \frac{Time(t) - Time(t_{start})}{Time(t_{end}) - Time(t_{start})})^p \| \right)^{\frac{1}{p}}$$

Because estimator selection operates at the level of individual pipelines (the weighted sum of which is the overall query progress) we report the error on the level of individual pipelines in the experiments.

### 6.1 Sensitivity Analysis for Estimator Selection

As a first step towards understanding how well our estimator selection component generalizes when the training and test queries vary in similarity, we systematically vary certain key parameters between the training and test workloads in this section. Because this setup requires many different instances of a number of different query templates, we use the TPC-H workload here.

In this study we ran a similar experiment to the example described in Section 1, evaluating the robustness of 3 major estimators – DNE, TGN and LUO – proposed in prior work as well as the estimator selection technique choosing among them.

**Evaluation Metric:** Something we found to hold throughout these experiments was that the overall average estimation error for the estimator selection was consistently lower than the one for the 3 individual estimators, despite the differences between training and test data. Hence, in the following, we will instead concentrate of a different measure, using the fraction of queries for with the estimator selection correctly predicts the optimal estimator to assess the sensitivity to the systematic changes.

**Varying the selectivity:** In this experiment, we study the effects of variation in the cardinality/selectivity on estimator selection. Here, we only consider operator pipelines that occur at least 6 times in total in the workload; for each pipeline, we sort all instances in the workload by their total number of GetNext calls and then group them into 3 equal-sized "buckets", the first one containing the pipelines with the smallest overall number of GetNext calls, the last group the pipelines with the highest number of GetNext calls and the remainder in the middle group. We then perform 3 experiments, in each of which we use two of the groups for training and the other one for testing. In Table 2 we show the percentage of pipelines for which each individual estimator has the lowest error as well as the percentage of pipelines for which the estimator selection model correctly selects this estimator.

As we can see, the percentages at which different estimators are optimal vary significantly between the different test sets (e.g., in case of TGN, the difference is more than 30% between the 1st and 3rd experiment). One result of this variance is that in the first experiment the estimator selection module chooses the optimal estimator at a lower rate than choosing TGN for every query would result in. While the resulting average progress estimation error is still the smallest for estimation selection (and not for TGN), meaning that estimator selection still is more robust in this scenario than any individual estimator, this indicates estimation selection is somewhat sensitive to large differences in cardinality, meaning that the training data for any estimator selection model should ideally cover a wide range of cardinalities.

| Test Set: | "small" data | "medium" data | "large" data |
|---|---|---|---|
| **Estimator** | % optimal | % optimal | % optimal |
| DNE | 54.2% | 64.2% | 57.9% |
| TGN | 28.7% | 28.1% | 24.9% |
| LUO | 17.0% | 7.7% | 17.2% |
| EST. SEL. | 57.0% | 46.3% | 57.2% |

**Table 5: Sensitivity Analysis: varying the data size.**

**Varying the physical design:** Next, we use a similar setup to before, but vary the physical database design, using the three different physical designs for TPC-H described in the beginning of Section 6, corresponding to different levels of tuning. As shown before in Table 1, these changes in physical design significantly impact the operators used in the execution plans and hence the pipelines seen in progress estimation. The resulting error-ratios are shown in Table 3 – with the test set being the "*untuned*" database in the right column, "*partially-tuned*" in the middle one and "*fully tuned*" in the leftmost one. Here, we see estimator selection perform extremely well (and significantly better than any individual estimator); in part this is due to the fact that we see smaller variance in the rates at which individual estimators are optimal across the different experiments. Estimator selection performs least well for the "*fully tuned*" test data; from Table 1 we can see that this workload has a very different operator mix than the other two, being comprised of many more nested iterations and containing more batch sorts, for which estimator selection becomes more difficult due to potential variance in the number of GetNext calls for each Driver Node tuple. However, our approach appears to be robust to the types of physical design changes seen in this experiment.

**Varying the data skew:** For this experiment, we generated the TPC-H database with several different skew factors $z = 0, 1, 2$ [1]. For each of these databases, we re-ran the 1000-query TPC-H benchmark, generating three sets of data; for each of the experiments, we then used two of the data sets for training and queries from the remaining one for testing. The resulting selection accuracy is shown in Table 4. Varying the skew factors introduces very significant changes in the workload – query plans vary significantly between skewed and non-skewed data and the (on average) best single estimator varies across these workloads as well: for skewed data DNE outperforms TGN and LUO, for $z = 0$ TGN outperforms the other two. This in turn makes the task of learning the estimator selection module very challenging, as the underlying biases are very different, making this experiment a serious test of our ability to generalize. Still, we can see that the estimator selection performs significantly better than any individual estimator, even though the overall accuracy at which we can select the optimal estimator is reduced when compared to the two previous experiments.

**Varying the data size:** In this experiment, we use different scale-factors of the underlying TPC-H database in the training/test set, using three different TPC-H databases with scale-factors 2, 5 and 10. Interestingly, and similarly to the case of variations in data skew and physical design, these changes lead to rather different query plans, which in turn challenge our technique's ability to generalize (as some pipelines in the test data may not have been seen in training). The resulting selection accuracy is shown in Table 5, with the test data going from small (2GB) to large (10GB) from the left to the right column. Here, our technique performs better then the individual estimators for one experiment, but only even to DNE for $Z = 2$ and at a rate worse than DNE for $Z = 1$, making this setup the most challenging for our technique's ability to generalize. This is also reflected in the fact that the rates at which estimator selection selects the optimal estimator is reduced when compared to the previous experiments.

Overall, the estimator selection approach is mostly able to generalize well across all of the different variations in workload parameters we studied in this section. For each of the experiments, the average estimation error is lower than it would be for any individual estimator; however, significant changes in the data skew and size between training and test data noticeably reduce the rate at which our technique is able to select "correctly". Consequently, a range of data sizes and skews should be used as part of model training. We will study the performance of estimator selection in the context of fully "ad-hoc" queries next.

## 6.2 Estimator Selection for "Ad-Hoc" Queries

In this experiment, we are interested in evaluating the robustness of estimator selection in the case of fully "ad-hoc" queries, where there is *no overlap* between training and test workloads.

**Test/Training setup:** We use the 6 workloads described earlier for this experiment, but ensure that in the experiments each of the 6 workloads is either only part of the training set that we use to train the estimator selection module, or the test set that we use to evaluate it, but never both. Each experiment is based on a test set containing all queries from one of the workloads and a training sets consisting of the other 5, resulting in a total of 6 test/training combinations. All aggregate statistics are now computed by averaging over the results of each of the corresponding 6 experiments.

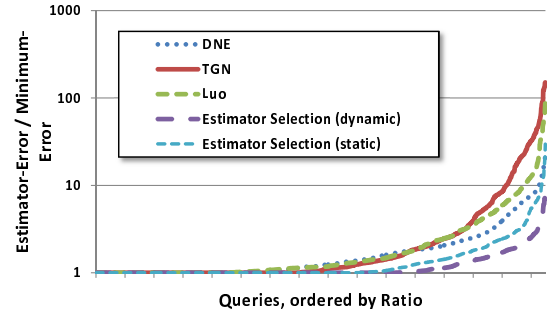

**Figure 4:** Compared to the DNE, TGN and LUO estimators in isolation, estimator selection improves robustness significantly. Considering dynamic features improves the estimation error further.

We first compute the ratios between the optimal error and the one of each estimator (as well as estimator selection), similarly to the experiments in Section 1; the resulting graph is shown in Figure 4. We can see that using the estimator selection module results in the progress estimator with the smallest estimation error being selected for a much larger fraction of the queries than would be the case otherwise – whereas using DNE, TGN and LUO is optimal for 31%, 44% and 25% of the queries, estimator selection selects the optimal estimator for 55% (static features) or 64% (dynamic features) of the queries. More importantly, the error (relative to the optimal estimator) is now significantly less in the cases where the best estimator is not selected.

To quantify this, Table 6 displays the percentage of cases where the ratio of estimation error to minimum error exceeds 2x, 5x and 10x for the different techniques. As we can see, estimator selection is significantly more robust than the alternatives – when using dynamic features, less than 1% of all queries have a ratio of more than 5x, whereas this is the case for 7.8%–14.5% of the queries for the three estimators in isolation.

|     | DNE   | TGN   | LUO   | EST. SEL. (ST) | EST. SEL. (DY) |
|-----|-------|-------|-------|----------------|----------------|
| 2x  | 23.6% | 26.7% | 27.3% | 13.2%          | 6.3%           |
| 5x  | 7.8%  | 14.5% | 11.4% | 3.7%           | 0.8%           |
| 10x | 1.6%  | 8.9%  | 5.0%  | 1.0%           | 0.3%           |

**Table 6:** Percentage of pipelines where the ratio of estimation error to minimum error is larger than 2x, 5x and 10x.

To quantify the effects of estimator selection on the actual progress estimation error, we computed the average $L_1$ and $L_2$ error when using the DNE, TGN and LUO estimators exclusively as well as the error resulting from using estimator selection. For this estimator selection experiment, we evaluated the case where only the three original estimators can be selected, as well as the case in which the novel estimators BATCHDNE, DNESEEK and TGNINT are also available. We evaluated both cases when using static features only as well as using static + dynamic features. The results are plotted in Figure 5. As we can see, estimator selection results in significant improvements in accuracy over the originally proposed estimators, with the combination of static and dynamic features outperforming static features alone, especially when only the three initial progress estimators where available to the selection module. When additional estimators are added, the error is reduced further and the difference between the two feature sets is much less pronounced. Note that while the differences between absolute errors in this figure do not appear to be very large, this is partially due to the fact that the distribution of errors across pipelines is somewhat "skewed": a significant fraction of the pipelines are "easy", with most progress estimators showing small errors when used on them. In contrast, a significant fraction of pipelines is "hard" in the sense that nearly all (but not all) of the progress estimation techniques exhibit large errors. It is for this subset of pipelines that estimator selection techniques are crucial.

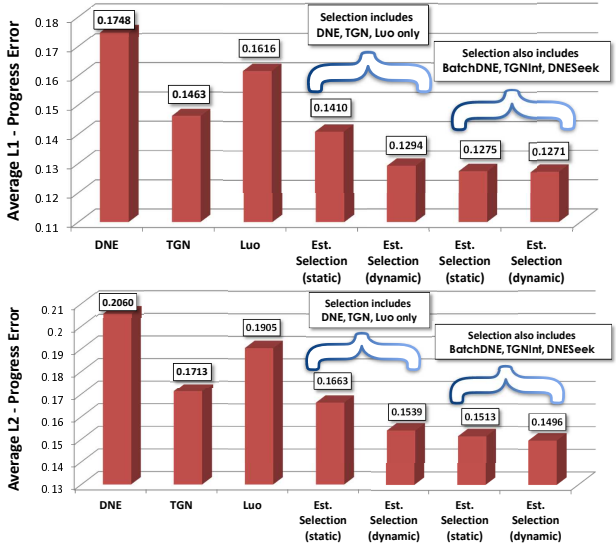

**Figure 5:** Estimator selection improves robustness significantly, especially as more estimators are added.

To evaluate the quality of the estimator selection itself (i.e., independent of the quality of the underlying estimators), we also computed the errors resulting for a theoretical optimal "oracle" estimator selection module which always picks the progress estimator with the smallest error. For the scenario where the available estimators are DNE, TGN and LUO, the resulting $L_1$ error was 10.9, whereas when BATCHDNE, TGNINT and DNESEEK are added, the $L_1$ error becomes 9.9. This suggests that estimator selection is already very accurate (close to the lower bound of the "oracle"), and that the main cause of the remaining error are queries for which none of the available estimators are accurate to begin with.

**Worst-case Optimal Estimators:** We also evaluated both the SAFE and PMAX estimators, but did not plot their errors, as both of them had more than 2x the error of the worst-performing alternative estimator in practice – here, PMAX performed worse, with an $L_1$ error of 0.50 ($L_2$ error 0.58), whereas SAFE had an $L_1$ error of 0.40 ($L_2$ error 0.45). These results appear to rule both of them out for practical applications.

### 6.3 Error Analysis

In order to understand the nature of the progress estimation errors we observed, we manually analyzed the differences between the estimated and true progress for a large number of different pipelines in our experiments; examples can be seen in Figures 6 and 7. We observed that the errors between true and estimated progress are typically not distributed evenly across a query's duration, but that the points of largest absolute deviation typically occur close to the end of a query (whereas estimation is more accurate near the beginning). This puts the relative errors seen in Figure 5 in perspective, as the errors shown there average over all observations in a query and the differences in the averages are much smaller than the differences in the largest absolute deviations (for different estimators) we saw. Since long-running pipelines often (e.g., in the two real workloads used in the experiments) execute for hours, these deviations can correspond to intervals of 30 minutes or more.

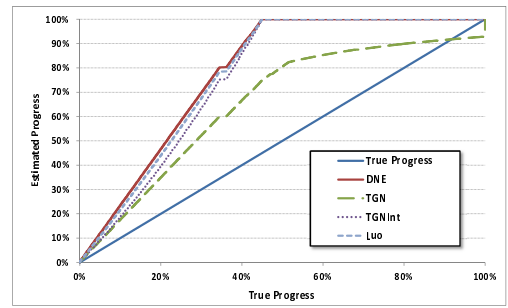

**Figure 6:** Example of progress estimation errors in a nested loop join pipeline: here, the partially blocking batch sort operations may result in significant errors for estimators heavily based on *Driver Nodes*.

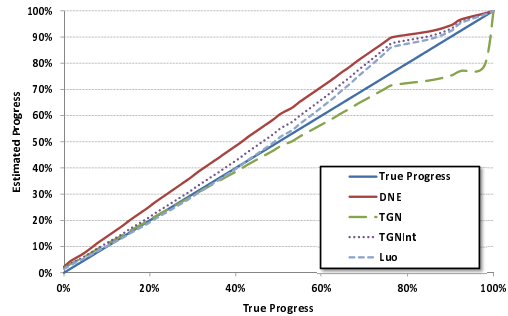

**Figure 7:** Example of progress estimation errors in a complex hash join query: selectivity estimation errors may affect the TGN estimator more, as it doesn't use interpolation (like LUO or TGNINT) or leverage *Driver Nodes* (like DNE or LUO).

Concerning the errors for different types of pipelines, we also observed that the sources of errors in them are fundamentally different; in the case of nested iterations such as in Figure 6, the observed errors often stem from inaccurate cardinality estimates, combined with the fact that the bounding/interpolation techniques described in Section 3.3 do not work for these cases, as they offer no meaningful bounds on the number of joining tuples on the inner side of a join.

Moreover, the introduction of bach sort operations leads to situations in which all GetNext calls at the driver nodes have completed,

but the pipeline is still far from finished, leading to a severe overestimate of the progress for techniques using driver nodes heavily (such as DNE, or LUO); issues like this have lead to the formulation of the additional progress estimators described earlier, such as the BATCHDNE estimator (Section 5.1).

For pipelines without nested iteration, estimators leveraging the fraction of consumed driver node input can adjust their estimates later during the query execution (as seen in Figure 7). In contrast, TGN may provide initially better estimates due to taking into account work done at intermediate nodes in the query plan, but does not have a way to recover from cardinality errors. Refining estimator selection based on dynamic features can thus offer performance superior to any of the individual estimators in this scenario, e.g., selecting TGN at the start of the query and LUO near the end.

### 6.4 Training Times

The key parameters that determine the overhead of training the regression models are the number of boosting iterations $M$ and the number of training examples (training times are independent of data volume or query execution times). Table 7 shows the training times (in seconds) for various parameters. These times include the time taken for reading in the training data and writing the output model to disk. As we can see, the training cost is very small, even for very large training sets. Note here that training data can be captured at low overhead in a running system – all we need to do is to compute the overall estimation error for the different estimators for each query which we want to add to the training set; since all estimators are essentially based of the same small set of counters, the overhead for tracking multiple estimators is nearly identical to the overhead for computing a single one. We then write out these errors and all features for a query (corresponding to about 200 double values); as progress estimation is mainly of interest for long-running queries consuming significant resources, this overhead is typically not significant. Combined with the low overhead for model training, this means that it is possible to further adapt the estimator selection component in running system by capturing estimator behavior and retraining the underlying models without incurring significant cost.

| Training | Boosting Iterations (M) | | | | | |
| --- | --- | --- | --- | --- | --- | --- |
| Examples | 20 | 50 | 100 | 200 | 500 | 1K |
| 100 | < 1 | < 1 | < 1 | < 1 | < 1 | < 1 |
| 500 | < 1 | < 1 | < 1 | < 1 | < 1 | < 1 |
| 3K | < 1 | < 1 | < 1 | < 1 | 1 | 2 |
| 6K | < 1 | < 1 | 1 | 1 | 2 | 4 |
| 60K | 8 | 9 | 11 | 15 | 15 | 41 |

**Table 7: Training Times in seconds.**

### 6.5 Feature Importance

In this section, we study the relative importance of the various features proposed in Sections 4.3 and 4.4. For this purpose, we initially ran the following greedy feature selection method: first, we iterate through all proposed features and select the one that – when building the regression models described in Section 4.1 using only this feature as input – results in the smallest overall mean square error. Once we have identified this feature, we add it to the set of *selected features* $\mathcal{S}$. We then repeat the greedy selection process, with the only difference being that all features in $\mathcal{S}$ selected so far are also used for model construction. This way, we incrementally select the feature that – given a set of features selected so far – gives us the largest incremental gain in accuracy.

Here, the first feature selected was $SelBelow_{NL\_Join}$, which quantifies the number tuples in inputs to nested loop operators (relative to the overall number of tuples in the pipeline) and hence gives some indication of the potential variance in the "per-tuple work" for the driver-node input. The next feature selected was $Cor_{DNESEEK,4,20}$, which quantifies how well the DNESEEK estimator (which performed best overall among all individual progress estimators) correlates with time. Overall, the fact that these features are selected first demonstrates the critical nature of nested iteration for progress estimation (which has also been discussed in earlier work). The 3rd feature selected was $SelAt_{DN}$ which quantifies the fraction of all GetNext calls performed at the driver nodes themselves; a high value of this feature is a good indication of a query for which DNE is the optimal estimator.

Among the 10 features selected next, seven were dynamic features (six of them quantifying the correlation between estimator output and time); this is not surprising, as these features are the only way to incorporate the passage of time, when it is not captured through the *total GetNext* model.

### 6.6 How many Estimators do we need?

Given the improvements in accuracy we saw when using the additional estimators introduced in Section 5, one interesting question is to revisit the issue of single-estimator robustness we discussed in Section 1: if one of the new estimators is (close to) optimal for a sufficiently large fraction of queries, it may serve as a "default estimator" for all cases, eliminating the need for estimator selection. To evaluate this question, we computed – for each estimator – the fraction of times (over all workloads defined earlier) where it was "almost optimal", meaning that it was (a) either the optimal estimator, (b) the *absolute* difference between it and the optimal estimator was very small (less than 0.01) or (c) the *relative* difference between it and the optimal estimator was small (less than 1%).

The results of this experiment are shown in the first column of Table 8. Note that because for any give pipeline several estimators can be "almost optimal", the fractions in the table add up to more than 100%.

| Estimator | % (close to) optimal | % significantly outperforms |
| --- | --- | --- |
| DNE | 37.6% | 0.2% |
| TGN | 37.7% | 17.7% |
| LUO | 30.3% | 3.86% |
| PMAX | 0.2% | 0.06% |
| SAFE | 4.7% | 4.2% |
| BATCHDNE | 39.2% | 2.2% |
| DNESEEK | 45.5% | 9.4% |
| TGNINT | 31.1% | 6.69% |

**Table 8: Fraction of queries for which (a) estimators were (nearly) optimal or (b) significantly outperform *all* others.**

As we can see, neither of the new estimators is optimal for a significantly larger fraction of pipelines than the original ones and none of them even passes the 50% mark. The two highest-scoring estimators are BATCHDNE and DNESEEK, which improve upon DNE for specific sub-cases of nested-loop iterations, but do not extend beyond them. Thus, it appears we cannot rely on a single estimator for all queries. This again underscores the importance of accurate estimator selection techniques.

Using a similar methodology, we might ask how many (and which) estimators we require in estimator selection? If an estimator does not noticeably outperform *all—/* others on at least part of the space of SQL queries then we need not consider it in estimator selection. In the 3rd column of Table 8 we show for each progress estimator the fraction of queries for which it significantly outperforms all others (meaning that it (a) has the lowest error, (b) the *absolute* difference between it and the next-best estimator is larger than 0.01 *and* (c) the *relative* difference between it and the optimal estimator is larger than 1%).

As we can see, only two of the estimators (DNE and PMAX) do not significantly outperform the remainder for at least 2% of the instances. The main reasons for the small percentage for DNE is that two other estimators – BATCHDNE and DNESEEK – produce identical estimates to DNE for a significant fraction of instances (namely those not containing *Batch Sort* or *Seek* operators). All other estimators significantly outperform the alternatives in a sufficiently large fraction of cases to be considered necessary for estimator selection.

## 6.7 Validating the Total GetNext and Bytes Processed Models

One implicit assumption in all hardness results of [5] on progress estimation is that the theoretical *GetNext Model* of progress is in fact a good measure of the time taken by a query (which is the metric users typically care about). As a consequence, we evaluated the error seen when using this model, which corresponds to the TGN estimator described in equation (3), but using the true number of total GetNext calls $N_i$ (which we obtained after query execution) in place of their estimates $E_i$.

Similarly, we also evaluated the model proposed in [13] based on the number of bytes processed at the dominant nodes and the end of each pipeline, again while substituting the correct number of total bytes written/output in the estimator.

The main goal for this experiment was to assess if these idealized models do indeed capture the progress of a large body of queries/pipelines accurately or if these models have themselves need to be re-thought. For evaluation, we again use the same setup as in Figure 5; here, the *GetNext Model* of progress gives an $L_1$ error of 0.062 ($L_2$ error of 0.073). Note that this error is significantly lower than the one of the other individual estimators or the estimator selection, neither of which does have access to accurate cardinality information, however. The model of [13] (with accurate cardinalities) performs significantly worse, coming in with a $L_1$ error of 0.12 ($L_2$ error of 0.142), meaning that even with exact cardinalities, the model of [13] performs on-par with estimator selection.

In conclusion, the *GetNext Model* of progress appears to be a sound basis for theoretical modeling of progress estimation, given that it correlates well with execution time in practice. This model has the "advantage" of 100% accurate knowledge of cardinalities, which is not attainable in practice; however, the experiments above indicate that significant improvements for progress estimation may be possible by improving upon the current techniques used to refine cardinality estimates during query processing.

## 7. CONCLUSION AND OUTLOOK

In this paper we studied the problem accurate progress estimation, proposing a statistical model for selecting among a set of progress estimators for a given query. In particular, we are interested in the robustness of the proposed techniques, meaning that they should result in accurate prediction even if the queries/data used in training are different from the ones the models are deployed on. We identified a number of features that are weakly predictive of the different estimator errors and formulated statistical models based on them, which form the basis of the estimator selection. We found that these features generalize well across different queries and databases; thus, the estimator selection performs well even when faced with novel "ad-hoc" queries.

Given that the main source of errors in the current framework appears to be the quality of the underlying estimators and not the selection itself (see Section 6.2) one key to improving progress estimation further appears to be the study of additional progress estimators; when combining them with an estimator-selection framework, these may be somewhat specialized and address only a subclass of queries. Further, given the high accuracy of the theoretical *Total GetNext* model (see Section 6.7), which assumes accurate knowledge of cardinalities, a further venue towards improved progress estimation may be the study of better *online* cardinality refinement.

## 8. REFERENCES
[1] Program for TPC-H data generation with Skew. ftp://ftp.research.microsoft.com/users/viveknar/TPCDSkew/.
[2] TPC-H and TPC-DS Benchmarks. http://www.tpc.org.
[3] S. Agrawal, S. Chaudhuri, L. Kollar, A. Marathe, V. Narasayya, and M. Syamala. Database Tuning Advisor for Microsoft SQL Server 2005. In *VLDB*, pages 1110–1121, 2004.
[4] C. Burges, T. Shaked, E. Renshaw, A. Lazier, M. Deeds, N. Hamilton, and G. Hullender. Learning to Rank using Gradient Descent. In *ICML*, 2005.
[5] S. Chaudhuri, R. Kaushik, and R. Ramamurthy. When can We Trust Progress Estimators For SQL Queries. In *ACM SIGMOD*, pages 575–586, 2005.
[6] S. Chaudhuri, V. Narasayya, and R. Ramamurthy. Estimating Progress of Execution for SQL Queries. In *ACM SIGMOD*, pages 803–814, 2004.
[7] D. J. DeWitt, J. F. Naughton, and J. Burger. Nested Loops Revisited. In *PDIS*, pages 230–242, 1993.
[8] J. Duggan, U. Cetintemel, O. Papaemmanouil, and E. Upfal. Performance Prediction for Concurrent Database Workloads. In *ACM SIGMOD*, 2011.
[9] M. Elhemali, C. A. Galindo-Legaria, T. Grabs, and M. M. Joshi. Execution Strategies for SQL Subqueries. In *ACM SIGMOD*, pages 993–1004, 2007.
[10] J. Friedman. Greedy Function Approximation: a Gradient Boosting Machine. *Annals of Statistics*, 29(5), 2001.
[11] A. Ganapathi, H. Kuno, U. Dayal, J. L. Wiener, A. Fox, M. Jordan, and D. Patterson. Predicting Multiple Metrics for Queries: Better Decisions Enabled by Machine Learning. In *IEEE ICDE*, pages 592–603, 2009.
[12] G. Luo, J. Naughton, and P. Yu. Multi-query SQL Progress Indicators. In *EDBT*, pages 921–941, 2006.
[13] G. Luo, J. F. Naughton, C. J. Ellmann, and M. W. Watzke. Toward a Progress Indicator for Database Queries. In *ACM SIGMOD*, pages 791–802, 2004.
[14] G. Luo, J. F. Naughton, C. J. Ellmann, and M. W. Watzke. Increasing the Accuracy and Coverage of SQL Progress Indicators. In *IEEE ICDE*, pages 853–864, 2005.
[15] C. Mishra and N. Koudas. The Design of a Query Monitoring System. *ACM Trans. Database Syst.*, 34:1:1–1:51, April 2009.
[16] C. Mishra and M. Volkovs. ConEx: a System for Monitoring Queries. In *ACM SIGMOD*, pages 1076–1078, 2007.
[17] K. Morton, M. Balazinska, and D. Grossman. Paratimer: A Progress Indicator for MapReduce DAGs. In *ACM SIGMOD*, pages 507–518, 2010.
[18] F. Sebastiani. Machine Learning in Automated Text Categorization. *ACM Computing Surveys*, 34(1):1–47, 2002.
[19] Q. Wu, C. J. Burges, K. M. Svore, and J. Gao. Ranking, Boosting, and Model Adaptation. Technical report, Microsoft Research, 2008.